\documentclass[11pt,twoside,letterpaper]{article} %% The same for {book}
\usepackage{times,fancyhdr,deluxetable}
\usepackage[dvips]{graphicx}
\makeatletter 
\def\cleardoublepage{\clearpage\if@twoside \ifodd\c@page\else% 
    \hbox{}% 
    \thispagestyle{empty}%
    \newpage% 
    \if@twocolumn\hbox{}\newpage\fi\fi\fi} 
\makeatother 
\def\figurename{Figure}
\makeatletter
\renewcommand{\fnum@figure}[1]{\figurename~\thefigure.}
\makeatother

\def\tablename{Table}
\makeatletter
\renewcommand{\fnum@table}[1]{\tablename~\thetable.}
\makeatother

\begin{document}
\title{
{\begin{flushleft}
\vskip 0.45in
{\normalsize\bfseries\textit{Chapter~1}}
\end{flushleft}
\vskip 0.45in
%
%
%%%%%%%%%%%%%%%%%%%%%%%%%%%%%%%%%%%%%%%%%%%%%%%%%%%%%%%%
%
%
% AUTHOR:  This belongs to you 
%%%%%%%%%%%%%%%%%%
\bfseries\scshape Neutron starquakes and the dynamic crust}}
\author{\bfseries\itshape Anna L. Watts\thanks{E-mail A.L.Watts@uva.nl}\\
Astronomical Institute `Anton Pannekoek', University of Amsterdam \\
Science Park 904, 1090GE Amsterdam, The Netherlands}
\date{}
\maketitle
\thispagestyle{empty}
\setcounter{page}{1}
% ------- [First Page Running Head] - place it immediately after title! ------
\thispagestyle{fancy}
\fancyhead{}
\fancyhead[L]{In: Neutron Star Crust \\ 
Editors: C.A. Bertulani and J. Piekarewicz, pp. {\thepage-\pageref{lastpage-01}}} % needs \label{lastpage-01} on the last page.
\fancyhead[R]{ISBN 0000000000  \\
\copyright~2012 Nova Science Publishers, Inc.}
\fancyfoot{}
\renewcommand{\headrulewidth}{0pt}
\vspace{2in}
\noindent \textbf{PACS} 26.60.+c · 97.10.Sj · 97.60.Jd
\vspace{.08in} \noindent \textbf{Keywords:} Magnetars. Neutron stars. Seismology.
%
% ------------ [Running Heads - for odd and even pages] - please insert it only on page 2!
\pagestyle{fancy}
\fancyhead{}
\fancyhead[EC]{A.L.Watts}
\fancyhead[EL,OR]{\thepage}
\fancyhead[OC]{Neutron starquakes and the dynamic crust}
\fancyfoot{}
\renewcommand\headrulewidth{0.5pt} 
%------------------------------------------------------------------------------
%
\section{Introduction}

\begin{figure}[ht]
\begin{center}
\includegraphics[width=14cm, clip]{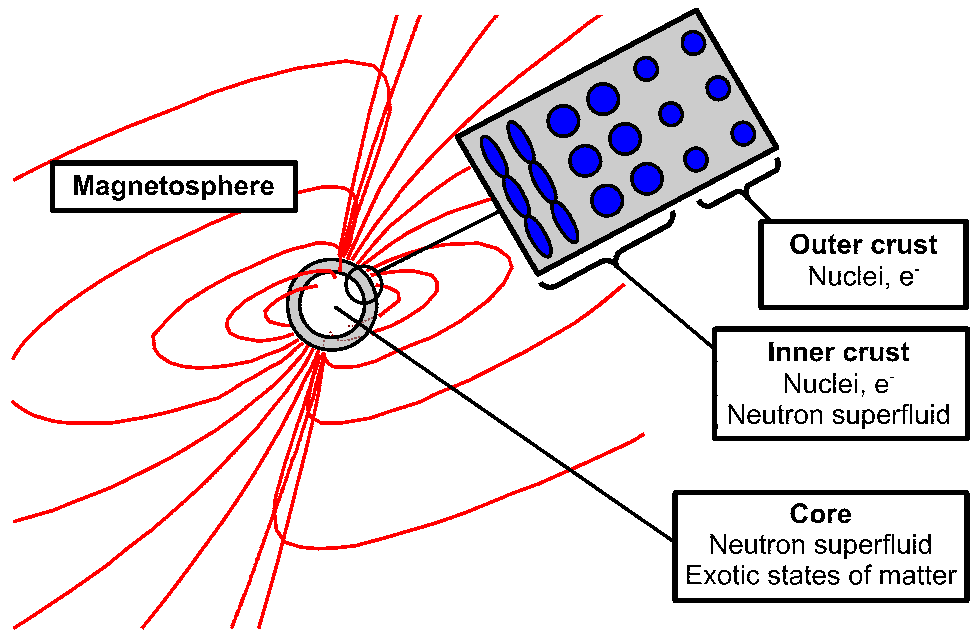}
\caption{$~$ Magnetic field lines (red, core field not shown) tangle as the field decays.  The field is locked to the lattice of charged nuclei in the neutron star crust, putting it under stress.  Stress can also build up in the magnetosphere, where low plasma density can hinder reconfiguration.  Catastrophic stress release -– a starquake –- allows field lines to reconfigure, generating flares. }
\label{cutfield}
\end{center}
\end{figure}

The most strongly magnetized neutron stars, the magnetars, have fields
that can be as high as $\sim 10^{15}$ G \cite{Woods06,Mereghetti08}.  Magnetars
are highly active, with spectacular outbursts of gamma-ray flares
powered by decay of the magnetic field.  The rapidly changing field is
strong enough that it should be able to stress and rupture the star's
crust (Figure \ref{cutfield}), with potentially interesting seismic consequences
\cite{Thompson95, Duncan98}. The existence of a dynamical relationship
was finally confirmed in dramatic fashion by the discovery of
long-lived seismic vibrations excited by rare giant flares
\cite{Israel05, Strohmayer05, Watts06, Strohmayer06}.  The starquakes associated with
giant flares are, it seems, so catastrophic that they leave the whole star
ringing. 

These discoveries have opened up the possibility of using
asteroseismology to study the extreme conditions of the neutron 
star interior.  Early work showed how magnetar vibrations could in
principle put extremely 
tight bounds on the composition of dense matter \cite{Strohmayer06, Samuelsson07,
  Lattimer07, Watts07} and interior magnetic field strength, which is
very hard to measure directly \cite{Israel05}.  This triggered a major
theoretical effort to improve models of global seismic oscillations
and include key physics.  Much of this work
has concentrated on the effects of the strong magnetic field, which
couples the solid crust and fluid core together, complicating mode
calculations.  The presence of superfluid neutrons is now known to
have a major impact, and we have uncovered serious uncertainties in
our understanding of the composition of the lattice nuclei in the
solid crust.

In Section \ref{gso} I review the main observational properties of the
oscillations 
seen in the giant flares, and in Section \ref{gsomodels} the state of
the art in terms of 
theoretical models of global seismic vibrations in magnetars.  In Section \ref{sf} I discuss efforts to extend oscillation searches to smaller flares (which
are more frequent but less energetic).  Finally, in Section
\ref{rupture} I discuss the role that crust rupturing might play in
triggering magnetar flares and exciting global seismic
oscillations.

\section{Quasi-periodic oscillations in giant flares}
\label{gso} 

\subsection{Giant flares from SGR 1806--20 and SGR 1900+14}

\begin{figure}[ht]
\begin{center}
\includegraphics[width=14cm, clip=true]{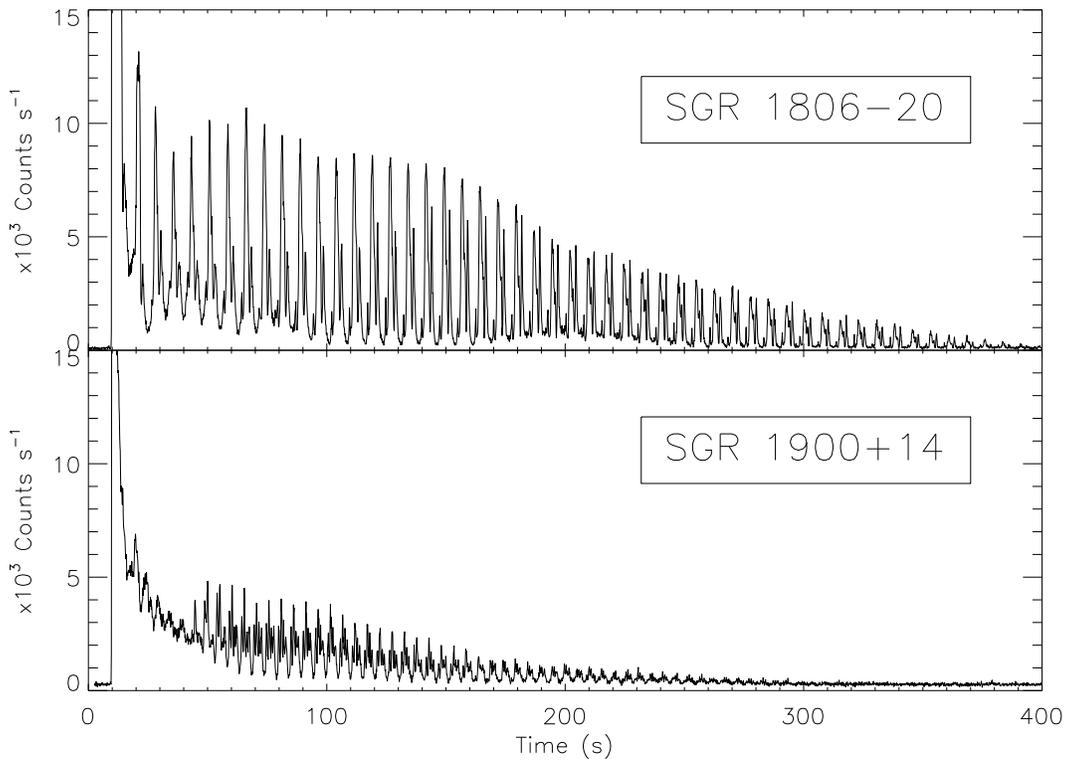}
\caption{$~$ RXTE PCA (2-90 keV) 0.125s resolution lightcurves for the 2004 giant flare from SGR 1806-20 and the 1998 giant flare from SGR 1900+14. Note that there were data
  gaps in the highest time resolution data stream for the SGR 1900+14
  event that reduced the amount of data available for high frequency
  QPO searches (shown here is the lower time resolution Standard1 data
  stream).  The PCA is completely saturated in the peak of the
  flares, despite the fact that neither event is on-axis for the
  telescope.  Thereafter strong pulsations can be seen at the spin
  period in both events (7.6s for SGR 1806--20 and 5.2s for SGR
  1900+14).  The rotational pulse profile is complex, with two main
  peaks for SGR
  1806-20 and four main peaks for SGR 1900+14.}
\label{lc}
\end{center}
\end{figure}

On December 27th 2004, the most energetic giant flare ever recorded
was detected from the magnetar SGR 1806--20 \cite{Terasawa05,
  Palmer05,Hurley05}.  Analysis of data from both the {\it Rossi
  X-ray Timing Explorer} (RXTE) Proportional Counter Array (PCA) and
the {\it Ramaty High Energy Solar Spectroscopic Imager} (RHESSI)
revealed a set of highly significant Quasi-Periodic Oscillations
(QPOs) in the decaying tail of the giant flare \cite{Israel05,
  Watts06, Strohmayer06}.  Subsequent re-analysis of RXTE PCA data from
the August 27th 1998 giant flare from the magnetar SGR 1900+14
\cite{Hurley99,Feroci99} 
revealed a similar set of QPOs \cite{Strohmayer05}.  Figure \ref{lc} shows the RXTE PCA
lightcurves of both giant flares.  

\begin{deluxetable}{ccccl}
\tabletypesize{\small} 
\tablecaption{Key properties of QPOs in giant flares.\label{qposum}} 
\tablewidth{0pt}
\tablehead{\colhead{Frequency (Hz)\tablenotemark{\star}} & \colhead{Width (Hz)\tablenotemark{\star}} &
\colhead{Amplitude (\% rms)\tablenotemark{\star}} & \colhead{Duration (s)\tablenotemark{\star}} & \colhead{Satellite}  }
\tablecolumns{5}
 \startdata
\sidehead{SGR 1806--20 giant flare in 2004 \cite{Israel05,Watts06,Strohmayer06}}
18 & 2 & 4 & 60--230 & RXTE, RHESSI \\

26 & 3 & 5 & 60--230 & RXTE,RHESSI  \\

30 & 4 & 21 & 190--260 &  RXTE \\

93\tablenotemark{\dag} & 2 & 10--20\tablenotemark{\dag} & 150--260 & RXTE, RHESSI \\

150  & 17 & 7 & 10--350 &  RXTE  \\

626 & 1--2\tablenotemark{\ddag} & 10--20\tablenotemark{\ddag} & 50--260\tablenotemark{\ddag} & RXTE, RHESSI \\ 

1837 & 5 & 18 & 230--245 &  RXTE  \\

\sidehead{SGR 1900+14 giant flare in 1998 \cite{Strohmayer05}}

28 & 2 & 4 & 55--110 &  RXTE  \\
53 & 5 & 7 & 55--110 &  RXTE  \\
84 & $\cdot\cdot\cdot$  & 10 & 55--60\tablenotemark{\ast} &  RXTE  \\
155 & 6  & 11 & 55--110 &  RXTE  \\
\enddata
\tablenotetext{\star}{The frequency is the centroid frequency from a
Lorentzian fit; the quoted width is the associated full width at half
maximum.  All amplitudes are the values computed from rotational phase
dependent power spectra. The duration is given with respect to the
main flare at time zero (see Figure \ref{lc}). The nominal energy band in which the
QPOs are seen is $< 100$ keV except where noted.}
\tablenotetext{\dag}{Appears to drift upwards in frequency and
  amplitude over time.}
\tablenotetext{\ddag}{Seen at early times in $100-200$ keV RHESSI data at high amplitude,
 then at later times in RXTE data with lower amplitude/coherence and
 at different rotational phase.}
\tablenotetext{\ast}{Signal actually appears to be concentrated in an even
  shorter, $\sim 1$ s, time window.}
\end{deluxetable}

The frequencies, and some other properties of the QPOs, are
summarized in Table \ref{qposum}.   There are several key
characteristics that need to be explained by any theoretical model.  

\subsection{Frequencies}
\label{freqs}

The frequencies of the observed QPOs lie in the range $\approx$ 18--1800 Hz.
Time resolution for both RXTE and RHESSI datasets was such that higher
frequencies could easily have been detected.  Sensitivity to lower
frequencies, however, is affected by the presence 
of a strong pulsation at the 
spin period and its
harmonics, as well as red noise from the overall decaying shape of the
giant flare
lightcurve. 

It is also clear that the signals are genuinely quasi, rather than
strictly, periodic.  The coherence $Q$ (defined as the ratio of
frequency to the width of the QPO) is however quite variable,
see Table \ref{qposum}.  One simple possibility that would explain the
quasi-periodicity is an exponentially decaying amplitude (as might
expected for oscillations triggered by a starquake).  An
exponentially decaying signal gives rise to a QPO with a width that is
related to the decay timescale.  By measuring width from the power
spectrum, one can therefore infer the decay timescale and hence calculate the time that would have
been necessary for the signal amplitude to fall, for example, to 5\%
of its initial level.  For all of the observed QPOs this time is very
short, $\sim 1$ s, far shorter than the signal duration that has
been inferred for most of the QPOs (see Section \ref{amps}).  One can therefore conclude that simple
exponential decay models cannot 
 explain the all of the QPO widths.   So what else might be
causing the broadening of the signal?  Options include frequency drift
and splitting (with closely-separated frequencies being unresolved).  Unfortunately existing data for the QPOs are not good
enough to resolve this question, but there is tentative evidence
from the strongest QPOs suggesting that both possibilities may be
occurring \cite{Israel05, Strohmayer06}.  

\subsection{Amplitudes and duration}
\label{amps}

The QPOs are all found to be strongly rotational phase dependent --
with amplitude apparently varying strongly over the rotational cycle.
However the dominant rotational phase is not the same for all of the
detected QPOs.  The amplitudes of the QPOs, in the rotational phase
where they are strongest, can be very high (see Table \ref{qposum}).
Where it can be measured, amplitude seems to increase with photon
energy.  However some caution must be exercised with energy-dependent
analysis since none of the events were on-axis for the detectors so
there would have been 
substantial scattering of photons within the body of the spacecraft
prior to their interaction with the detectors.  For this reason the
published analysis considers only very broad energy bands.    

For the strongest QPOs, which can be detected in relatively short time
windows, it is possible to measure duration reasonably precisely.  For
the weaker QPOs, where multiple cycles must be stacked to record a
detection, duration can be estimated by analysing how signal strength
varies once extra cycles are introduced to the analysis.  From this we
infer that many of the QPOs persist for much of the decaying tail,
with some only becoming visible some time after the main peak of the
flare.  By contrast, other QPOs seem to be detectable in only very
short time windows (Table \ref{qposum}).  Pinpointing the precise
point at which signals appear and disappear in the current data is
however limited by both statistics and the difficulty of tracking
variable-amplitude signals that drift in frequency across rotational cycles.

\subsection{Additional analysis}

Since the initial QPO discovery papers
\cite{Israel05,Strohmayer05,Watts06,Strohmayer06}, some additional
analysis of the QPOs in the giant flares has been published.  A more
thorough rotational-phase and photon energy-dependent search for QPOs
in the tail of the giant flare from SGR 1900+14, motivated by a search
for QPOs with a frequency $\sim 10$ Hz, revealed no additional
significant signals \cite{Steiner09}.   However the upper limits that it was possible
to set on amplitude were still higher than the amplitude of the lowest frequency (18 Hz) QPO detected in the tail of the
giant flare from SGR 1806--20, due to the lower quality of the data
available.

Recently \cite{Hambaryan11} published the results of a new analysis of
the RXTE data from
the tail of the SGR 1806--20 giant flare.  There were several differences in analysis technique compared to the earlier results of
\cite{Israel05,Strohmayer06}.  Firstly the authors took more rigorous
account of the effects of the background lightcurve variations in
assessing the 
significances of the detections (see discussion in Section \ref{sf}),
confirming the existing detections.  They used a 
Bayesian method to assess statistical significances rather than the more
traditional methods used in the previous papers.  They also conducted
searches for shorter duration QPOs, finding several additional
frequencies in the range 17--117 Hz.  The previous analysis, by contrast, focused more on long
duration signals (the duration of intervals to be searched is one of
the choices that must be made during the analysis).  It would be
interesting to see whether these short duration signals are seen,
with comparable significances, using the more traditional
method -- assuming that the choices made about the number of different
time intervals to search can be properly factored into the numbers of
trials.  

Gravitational wave astronomers have also joined the hunt for QPOs in
giant flares.  The LIGO Hanford detector was taking science data at
the time of the SGR 1806--20 giant flare in 2004.  Analysis of the
data stream for signals at the same frequencies as those seen in the
X-ray and Gamma-ray data reported no significant detections, but were
able to place upper limits on the strength of any associated
gravitational wave emission at these frequencies \cite{Abbott07}.
Although the limits were not particularly constraining for models,
this bodes well for future giant flares, as
gravitational wave detector sensitivity continues to improve.

\subsection{Impulsive phase variability}

There is also some evidence for variability in the earlier, impulsive
phase of the giant flares.  In an analysis of Prognoz and Venera data from
the first 200 ms of the 1979 giant flare from the magnetar SGR 0526-66
\cite{Mazets79} (the only high time resolution data available for this
event) \cite{Barat83} reported evidence for variability at a frequency of $\approx 43$
Hz.  Geotail observations of the first 500 ms of the SGR 1806-20 giant
flare indicated periodicity at $\approx 50$ Hz, but no similar
periodicities are seen in the SGR 1900+14 giant flare
\cite{Terasawa06}.  Analysis of variability in the impulsive phase of
giant flares is however complicated by effects such as dead time.

\section{Theoretical models for global seismic oscillations}
\label{gsomodels}

\subsection{Early models}

In 1998, well before the discovery of the QPOs, \cite{Duncan98} had
suggested that the starquakes associated with giant flares might excite global 
seismic oscillations (GSOs) of the neutron star. It was argued that the easiest
modes to excite would be toroidal (or {\it torsional}) shear modes of
the neutron star crust. These modes, which involve primarily horizontal
displacements, are restored by the shear modulus of the ions in the
solid crust. Toroidal modes, unlike modes that involve compression or
vertical displacements, have much lower energies of excitation and
longer damping times.  Using earlier work by \cite{Hansen80,
  McDermott88} as a guide, and applying a correction for gravitational
redshift, \cite{Duncan98} estimated frequencies for the expected
modes, finding a fundamental toroidal shear mode frequency
of $\approx$ 30 Hz.  

The fact that frequencies in this range were discovered in both of the
giant flares provided immediate support for the
interpretation of the QPOs in terms of GSOs\footnote{Other models, including purely magnetospheric
  oscillations or phenomena associated with a residual disk of
  fallback material from the supernova, were considered but discarded \cite{Watts07b}.  However see also the paper
  by \cite{Ma08} that considers the role of standing sausage mode
  oscillations in flux tubes in the magnetosphere.} \cite{Israel05}.  The GSO
model seemed to provide a natural explanation for both the observed
frequencies and the fact that 
similar phenomena were seen in giant flares from two different
magnetars.  It also opened up the possibility of using
asteroseismology to constrain the interior properties of neutron
stars.  

The torsional shear mode models of \cite{Duncan98} used a `free-slip'
boundary condition between the solid crust and the fluid core of the
star.  The earliest QPO papers showed that it was indeed possible
to fit many (although not all) of the observed frequencies
with sequences of such torsional shear modes, with the higher
frequencies corresponding to higher harmonics \cite{Israel05,
  Strohmayer05}.   The discovery 
of the 625 Hz QPO in the SGR 1806-20 giant flare \cite{Watts06,
  Strohmayer06}, with a frequency in the range expected for the first
radial overtone of the toroidal shear crust mode \cite{Piro05}, caused
particular excitement.  Since fundamental 
and first radial overtone depend in different ways on stellar
compactness, identification of both frequencies has the potential to
put very tight constraints on the dense matter equation of state
\cite{Strohmayer06, Lattimer07, Samuelsson07}.

The lowest QPO frequencies, however, did not fit within the framework
of the existing toroidal shear mode models, since their frequencies were
lower than the estimates of the fundamental toroidal shear mode
frequency. In their discovery paper,
\cite{Israel05} suggested that these lower frequency QPOs might be
torsional Alfv\'en modes 
of the fluid core.  If this interpretation were correct, the mode
frequencies could then be used to measure the interior magnetic field
strength. Although it is possible to estimate the dipole field
strength using spin-down measurements, the strength of the internal toroidal
component has yet to be measured directly for any
neutron star (although its magnitude can be estimated, for example, by
the energy budget required to power repeated magnetar flares,
\cite{Thompson95}).  Theorists have attempted to constrain the ratio
of poloidal to toroidal components by considering the stability of
magnetic equilibrium models  \cite{Braithwaite09, Lander09,Ciolfi09,
  Ciolfi11,Lasky11, Lander11a, Lander11b}. However the models still permit a range
of possibilities.  The results are also rather sensitive to choices of
boundary conditions, and as a consequence the models do not all agree.
An independent measure of internal field strength is essential.

Having established that GSOs were the most plausible model to explain
the QPOs, and their potential to explore both the dense matter equation of state
and the interior magnetic field, efforts then turned to improving the GSO
models.  The following sections summarize the main research
directions. 

\subsection{Magnetic field effects}
\label{magnetic}

Most of the work on GSO models has concentrated on the effects of the
strong magnetic field, which (as first pointed out by \cite{Levin06})
couples crust and core, thereby rendering the `free-slip' boundary
condition inappropriate.  This means that one has to consider global
magneto-elastic modes rather than treating the crust and core in
isolation.  A simple plane parallel slab model developed by
\cite{Glampedakis06} explored how this might work for a crust and core
coupled with a uniform magnetic field.  

A major complication in such a problem is that the Alfv\'en modes of a
magnetized fluid core (for a realistic, non-uniform, field
configuration) admit a continuum of solutions.  These can lead to unusual
time-dependent behaviours.  Work by \cite{Levin07} illustrated that
the presence of such a continuum could lead to the development of
drifting QPOs that might amplify when they passed close to the
`native' (uncoupled) frequencies of the crust or core.  This work also confirmed
that a perturbation applied to the crust would quickly excite motion
in the core.  

A number of groups have explored the problem of a magnetized crust that
remains decoupled from the fluid core, to understand how the native
crust frequencies would be affected.   Work by \cite{Piro05}, using a
simple slab model and a uniform magnetic field, showed that the radial
overtones of the shear modes were particularly susceptible to magnetic
modification for fields above $\sim 10^{15}$ G.  Studies using a
dipole field geometry in Newtonian gravity \cite{Lee07} and in GR
\cite{Sotani08b} confirmed this result, which was also found to hold
for more general field configurations \cite{Sotani08b}. These studies confirmed that strong magnetic
fields, of the order that one might expect in magnetars,
can indeed affect shear mode frequencies in the crust even before one
considers coupling to the core.  Another study explored the degree to
which magnetic splitting might affect mode frequencies
\cite{Shaisultanov09}.   

The nature of the Alfv\'en continuum was investigated initially using
models that treated only the fluid core of the star, without
considering the effect of the crust. A study by
\cite{Sotani08a} of torsional (axial) Alfv\'en oscillations of a star
with a dipole field, in GR, found two families of QPOs.  These could
be interpreted as the end points of different continua, one associated with
closed field line regions near the equator, and one with the open
field line regions near the magnetic pole.  A study of polar
oscillations for the same field geometry, by contrast, revealed only
discrete modes \cite{Sotani09}.  The authors pointed out that unlike
the axial oscillations, the polar
oscillations (which are restored by pressure as well as magnetic
forces in this problem) do not collapse to a zero-frequency spectrum in the zero-field
limit.  It is zero-frequency solutions which typically give rise to
continua.   They suggested that the presence of the crust in the
coupled problem, which would provide an additional restoring force,
might therefore remove the continua for the axial oscillations. Their
work was then extended by \cite{Colaiuda09}, who computed torsional
Alfv\'en modes of a fluid star for a background with a mixed poloidal/toroidal field
configuration.  This study also found two families of QPOs that seemed
to be associated with the end points of a continuum. This result was
confirmed by a fully non-linear study, also in GR
\cite{CerdaDuran09}.   Recent oscillation calculations in Newtonian
gravity by \cite{Lander10, Lander11b} have not found evidence for a
continuum for non-axisymmetric polar perturbations, for poloidal-only
and toroidal-only background field configurations.  Although the
studies are not directly comparable, this result seems to support the
conclusions of \cite{Sotani09} that continua may not affect polar
perturbations.   In summary, the presence of
continua seems to depend on both perturbation type and background field
configuration.  Torsional perturbations are however likely to be in
the affected category.  

Armed with a better understanding of the continuum, several groups are now
attempting the full magneto-elastic oscillation problem, taking
into account the coupling of crust and core.  Non-linear GR
simulations by \cite{Gabler11}, of axisymmetric torsional oscillations
of a star with a pure dipole field, continue to show two families of
QPOs that seem to be associated with Alfv\'en continua.  GR
calculations by \cite{Colaiuda11} for the same field geometry,
however, find a more complicated situation.  They recover
discrete crust-dominated and Alfv\'en
oscillations in addition to the continua.  This picture is backed up
by the work of \cite{vanHoven11}, who employ a simpler model to
explore the nature of the solutions to the coupled oscillator
problem.  These authors find gaps in the continuum frequency bands,
with strong
discrete gap modes developing if the `native' crust
frequencies happen to sit within these gaps.  These may be the discrete
modes found in an earlier Newtonian study of coupled crust-core
oscillations by \cite{Lee08}.   Drifting frequency behaviour, due to
the effects of the continua, seems to be
ubiquitous \cite{vanHoven11}.  

The studies that have been done on the effect of the magnetic
field are diverse, employing many different analytical and numerical
techniques.  One can nevertheless draw several clear conclusions from
this body of work.  Firstly, it has been confirmed that torsional modes can indeed be
easily excited by giant flares \cite{Levin11}.   An initial
perturbation applied to the crust will however excite motion in the
core on a very rapid ($<$ 0.1 s) timescale.  Crust and core therefore cannot be
considered in isolation, since the strong field couples the two
together, so GSO models must consider the global magneto-elastic oscillations
of the star.  The nature of such oscillations is now much clearer.  A realistic (non-uniform) magnetic field geometry admits
continuum solutions in various frequency bands.  The turning points
(end-points) of the continua can give rise to a strong response at
these frequencies (which can appear similar in a power spectrum to the
more usual discrete modes).  In addition, the presence of the continua
can lead to unusual behaviour including frequency drifting and
 amplitude variability.  Such dynamical responses are typical for continua (see for
example \cite{Watts03,Watts04}).  The system can also still admit
discrete modes, in the gaps between the various continuum frequency bands.  

The potential of the GSO technique is also still clear.  The
frequencies of the 
magneto-elastic oscillations depend on both the dense matter
equation of state and the interior field strength.  Crust physics, in
particular the shear properties, is still thought to be important in setting at least some of the
observed frequencies.  However no model can yet fit all of the
observations.  In addition, how GSOs modulate the X-ray emission and to give rise to
    properties such as the observed high QPO amplitudes remains an
    open question \cite{Timokhin08}.

\subsection{Superfluidity}
\label{superfluid}

The neutrons in the inner crust and core of the star are expected to
be in a superfluid state.  The superfluid in the inner core is
expected to affect the dynamics of the crust by virtue of the
interaction ({\it entrainment}) between the free neutrons and the lattice
nuclei.  The superfluid may also affect the dynamics of the core (and
hence the frequencies of Alfv\'en modes).  If the protons in the core
are in the form of a Type II superconductor, the magnetic field will
be carried in flux tubes, which can interact with the superfluid
vortices (the means by the which the superfluid supports rotation).
Work by \cite{vanHoven08} has shown that the Alfv\'en
modes of the core are not significantly mass-loaded by superfluid neutrons,
even if the vortices are strongly pinned to the flux tubes.  The
interaction between superfluid neutrons and the lattice nuclei in the
deep crust is more likely to be important.  Analysis of the
effect on global mode frequencies in the crust by \cite{Samuelsson09},
building on earlier local analysis by \cite{Andersson09}, suggests
that there might be at least a 10\% correction to the mode frequencies
due to superfluid effects. 

\subsection{Crust composition}
\label{crust}

Current results from the magneto-elastic simulations indicate that
the shear modulus of the crust is likely to set at least some of the
observed frequencies (Section \ref{magnetic}).  There is however a
substantial degree of uncertainty about these shear properties.  Particularly
problematic is the uncertainty in the density dependence of the
symmetry energy (the energy cost of creating an isospin asymmetry in
nucleonic matter - or in simpler terms, of creating matter with
unequal numbers of neutrons and protons) \cite{Steiner09}.  To obtain a fundamental
torsional
shear mode frequency $\approx$ 30 Hz, as computed by
\cite{Duncan98, Piro05}, requires a nuclear symmetry energy that depends only
very weakly on density.  If the nuclear symmetry
energy varies more strongly with density (a possibility well within
the current bounds of uncertainty) then the fundamental shear mode
frequency could be much lower, perhaps $\sim 10$ Hz.  This is due to
the effect on the composition of the nuclei in the crust lattice,
which in turn sets the shear modulus \cite{Ogata90, Strohmayer91}.
Lower crust frequencies would overlap
the expected range of the Alfv\'en continua and their turning points,
so could make identification of frequencies using coupled GSO models more
complex. 

GSO modelling may offer a way of testing strange star models.  Strange
stars are expected to have crusts with properties that are quite different
to those of normal neutron stars.  One possibility is for a
strange star to have a thin crust of normal nuclear material extending only
to neutron drip, suspended above a strange quark matter fluid core by
a strong electric field \cite{Alcock86}.  Another model posits a crust
in which nuggets of strange quark matter are embedded in a uniform
electron background \cite{Jaikumar06}.  Both types of crust are much thinner than
neutron star crusts, and have very different shear moduli.  A study by
\cite{Watts07} of the shear mode frequencies for such a crust has shown
that they are very dissimilar to those of neutron star crusts.  Fitting
the observed QPO frequencies was very difficult.  While that study did
not consider crust-core coupling, it seems unlikely that coupling
would change this conclusion since strange star crust shear mode frequencies
could fit neither low nor high frequency QPOs (so
magnetically-dominated oscillations
would have to account for all of the observations, something that seems
difficult, see Section \ref{magnetic}). However this should be
verified.   

\section{Oscillations in smaller bursts}
\label{sf}

During active phases magnetars emit numerous bright gamma-ray bursts
spanning orders of magnitude in duration ($10^{-2}-10^3$ s) and peak
luminosity ($10^{39}-10^{46}$ ergs/s), see for example \cite{Gogus99,Gogus00}.  The longest and most energetic
events (the giant flares) are extremely rare, only three having been
observed since the advent of high-energy (X-ray or gamma-ray)
telescopes.  This has motivated searches for lower level oscillations in the far more numerous smaller bursts.  Energetics of mode
excitation suggest that this is not unreasonable, and the smaller
flares clearly involve localized asymmetries that could be modulated
by seismic motion.  

A search for QPOs in a period of enhanced emission
with multiple bursts (a `burst storm'), from the magnetar SGR
J1550–-5418, has been carried out using data recorded by the {\it Fermi
  Gamma-ray Burst Monitor} (GBM) \cite{Kaneko10}.  The analysis
considered the entire period of the enhancement, both during and
between bursts.  No significant
signals were reported, with a $3\sigma$ upper limit on QPO amplitude  $\sim 10$\% rms
for frequencies in the range 100--4096 Hz.  Upper
limits for frequencies below 100 Hz are even less constraining.
Unpublished searches for QPOs from intermediate flares (such as the
August 29 1998 event from SGR 1900+14, \cite{Ibrahim01}) have also not
revealed any significant signals.  Techniques for
this kind of search are however being refined (see discussion of the
complications below) and further studies are ongoing.  Gravitational
wave searches for oscillations excited by regular magnetar bursts and
burst storms have not
made any detections, but have reported upper limits \cite{Abbott08,
  Abbott09, Abadie11}.

Recently, the detection of QPOs
in some short bursts from the magnetar SGR 1806--20 has been claimed
by \cite{ElMezeini10}.  There are however issues with the
analysis carried out by these authors that render the conclusions
doubtful.   The standard analysis technique, when searching for
periodic or 
quasi-periodic signals, is to use Fast Fourier Transforms to produce a
power spectrum (see \cite{vanderKlis89} for a comprehensive review). In the absence of any periodic
signal, the Poisson statistics of 
photon counting yield powers that are distributed as $\chi^2$
with two degrees of freedom.  When searching for 
quasi-periodic signals, it is often 
common to average powers from neighbouring frequency bins.   In
searches for weak 
signals one can also {\it stack}, or take an average, of power
spectra from many independent data segments (time bins) or bursts.  This affects
the number of degrees of freedom in 
the theoretical $\chi^2$ distribution of noise powers, but the modified
theoretical distribution is known.  

Having obtained a high power in a particular frequency bin, one first
 computes the probability of obtaining such a high power
through noise alone.  By this we mean the
chances of getting such a high value of the power, 
in the absence of a periodic signal, due to the natural fluctuations
in powers that are a by-product of photon counting statistics. One
does this using the known properties of the $\chi^2$ distribution
with the appropriate number of degrees of freedom.  One must then take
into account the number of trials, which depends on the number of
independent frequency bins, time bins, 
energy bands and bursts searched.  If the probability of such a power
arising through 
noise alone is below a certain threshold after taking into account
numbers of trials, then it is deemed significant, with the quoted
significance referring to the chances of having obtained such a signal
through noise alone. In the analysis of the SGR 1806--20 bursts,
\cite{ElMezeini10} appear to compute these numbers
correctly.  

A complication then comes from the fact that the noise powers are
not always in accordance with the theoretical
distribution.  The reason for this is simple.  What we are
searching for is a periodic signal {\it superimposed on
the overall rise and decay of a burst lightcurve}. This gives
low-frequency power and adds side-bands to noise powers, boosting
their level (for a nice discussion of these issues, see
\cite{Fox01}).  These problems are particularly acute in short bursts
when the background lightcurve is changing rapidly and there are sharp
edges that can lead to spurious high frequency variability.   

Assessing the true distribution of noise powers, that is to say the
values that they would take in the absence of a periodic signal, is
most commonly assessed using Monte Carlo simulations.  One makes a model
of the burst lightcurve, and then uses this as a basis to generate a
large sample of  
fake lightcurves with Poisson counting statistics but without any
periodic signals\footnote{Ideally effects such as deadtime and pileup
  should also be taken into account, see \cite{vanderKlis89}.}. 
Taking power spectra from these fake lightcurves then yields the true,
most likely 
frequency-dependent, distribution of noise powers. Taking into account
the varying 
nature of the background lightcurve should make the significance of
any claimed detection drop
compared to the significance computed from the ideal $\chi^2$
distribution.  Although \cite{ElMezeini10} do carry out simulations,
the significances that they quote rise substantially,
indicating a problem in their Monte Carlo simulation method.  Indeed
their simulated 
power spectra show far fewer high noise powers than one would expect given
the number of simulations carried out and the number of independent
frequency bins (enhancing the significance of any tentative
detection).  We conclude that the quoted significances are not robust,
and need to be revisited before the claimed detections can be verified.

\section{The role of crust rupturing}
\label{rupture}

The underlying cause of magnetar activity is decay of the ultra-strong magnetic field, which twists
the field lines into an unstable configuration \cite{Braithwaite06}. Once a tipping point is
reached, the field lines undergo rapid reconfiguration (perhaps involving reconnection, the
splicing of field lines). This creates currents whose dissipation generates gamma-rays. What is not
understood at all, however, is what triggers the flares. For flaring to be sporadic, there has to be
some barrier to magnetic reconfiguration that yields when a threshold is reached. The nature of this
barrier - and the trigger for the starquakes - is not known
\cite{Duncan04}.

One possibility is that reconfiguration is gated, or held up, by the crust of the star. The solid crust
can in principle resist motion as it is stressed by the changing interior field, yielding only when
magnetic force exceeds the breaking strain \cite{Thompson95, Perna11}. The resulting crust
rupture enables external field lines to move and reconfigure, generating the flare. Whether this is
the case depends on the breaking strain of the crust, set by its composition, crystalline structure, and
melting properties \cite{Horowitz09}. The strength of the deep crust, where the nuclei are
expected to be highly deformed (the so-called pasta phase), is likely
to be critical \cite{Pethick98}.

This is not the only option, however. Stress could also build up in the external magnetosphere,
being released only when plasma conditions permit reconnection via
various instabilities \cite{Lyutikov03, Gill10}. If the crust yields
plastically rather than resisting stress, then this may be more likely
\cite{Jones03}.  The excitation of GSOs then becomes an important
issue, however.  In addition it is now clear that flaring is not the
only way of transferring magnetic stress from the interior of the star
to the exterior.  The level of twist in the magnetosphere changes even
when there is no flaring, suggesting that non-violent stress transfer
is also possible \cite{Thompson02, Mereghetti05}.

\section{Conclusions}

The initial excitement associated with the discovery of the magnetar
QPOs was due in large part to the fact that they were thought to be
torsional shear modes of the neutron star crust.  The successful
identification of a sequence of modes, including radial overtones,
would have enabled a measurement of the thickness of the neutron star
crust and placed a very strong constraint on the dense matter equation
of state. 

The picture is, it now appears, more complex.  Crust physics
remains, nonetheless, an important part of the equation.  Global magneto-elastic seismic
oscillations remain the most plausible
explanation for the QPOs.  In these models the shear properties of the crust and restoring force
of the magnetic field couple together to create a complex oscillatory response
to the giant flare.  Mode identification in such a system will be more
challenging (and will benefit hugely from better data), but will be
far richer in terms of the revealed physics that initially imagined.

The role of the crust in gating magnetar flares also remains unclear.
How the crust yields under the influence of magnetic stress, in
particular whether this process is gradual or sudden, remains
unresolved.  Studies that are beginning to explore this issue are of
critical importance to understanding how and why magnetars burst.

\section{Acknowledgments}
My research is supported by a Netherlands Organisation for Scientific
Research (NWO) Vidi Grant. I would like to thank Daniela Huppenkothen,
Alaa Ibrahim, Gianluca Israel, Craig Markwardt and Michiel van der
Klis for discussions about the results of \cite{ElMezeini10}.

\label{lastpage-01}

\end{document}